\documentclass[journal,10pt,onecolumn,showkeys]{revtex4-1}
\usepackage{graphicx}
\usepackage{amsmath}
\usepackage{amssymb}
\usepackage{dcolumn}
\usepackage{latexsym}
\usepackage{rotating}
\usepackage[usenames,dvipsnames]{xcolor}
\usepackage{float}
\usepackage{epsfig}
\usepackage{psfrag}
\usepackage{natbib}
\usepackage{bm}
\usepackage{eucal}
\usepackage{braket}
\usepackage{enumerate}
\usepackage{longtable}
\usepackage{subfigure}
\usepackage{bm}
\usepackage{hyperref}
\usepackage{amsfonts}
\usepackage{dcolumn}% Align table columns on decimal point
\usepackage{bm}
\usepackage{subfigure}
\newcommand{\be}{\begin{equation}}
\newcommand{\ee}{\end{equation}}
\newcommand{\bn}{\begin{eqnarray}}
\newcommand{\en}{\end{eqnarray}}

\usepackage{color} % use colored text in latex

\usepackage{hyperref}
\begin{document}

\author{Chrislene Lionel, Shubham Das, Diparnab Banik, S. Koley$^{1}$}
\email{sudiptakoley20@gmail.com}

\title{Theoretical Study on Optoelectronic properties of Layered In$_2$O$_3$ and Ga$_2$O$_3$}
\affiliation{$^{1}$ Department of Physics, Amity Institute of Applied Sciences, Amity University,  
Kolkata, 700135, India}
\begin{abstract}
\noindent Composite oxides have been indeed proved to be valuable materials in 
optoelectronic applications. The combination of indium oxide and gallium oxide 
and other materials can lead to enhanced optical and electronic properties, making them suitable for a variety of optoelectronic devices. Meticulous 
analysis of the various optical properties helped to draw conclusions about 
the heterostructure of Indium and Gallium oxide and its use as a suitable
semiconducting material in the medium bandgap range. The density of states and 
	the band structure have been obtained from the density functional 
	theory calculations. Real frequency phonon density of states supports 
	dynamical stability of the crystal structure. A favorable energy band gap is achieved in the visible region of the spectrum, indicating that this mixed oxide is well suited for optoelectronic devices such as LEDs and solar cells.
\end{abstract}
%\pacs{
%71.10.Hf,
%Non Fermi Liquid
%}
\maketitle

\section{INTRODUCTION}
Nanomaterials have various electrical and optical properties that offer 
interesting applications in biomedicine (MRI contrast agents), cosmetics, 
batteries, textiles, paints, storage devices and optoelectronics 
(low-power LEDs)\cite{1}. Semiconductors usually have unusual band structures 
that allow them to be used in various applications. Following the discovery of 
new materials or the use of composite materials, great progress can be made in 
the area of optoelectronics. Nanomaterials enhance chemical reactivity of  
materials, which alters their strength. Oxides such as titanium dioxide and 
zinc oxide are 
transparent at the nanoscale, absorb and reflect ultraviolet rays and are 
therefore used in sunscreens\cite{2,3,4,5,6,7}. Iron oxide, a nanomaterial, 
helps as a pigment in lipsticks\cite{8}.

In recent studies conducted about semiconductor oxides, the doped oxides have 
shown significant and promising features, and applications in low-power 
electronics, memories, gas sensors, electrochromics and display applications\cite{wang}. 
Nickel oxide played a vital role in the development of resistive memories, 
as well as a hole-transporting layer in photovoltaic and optoelectronic devices.
Transparent p-n junctions also have been employed using p-type oxides with 
magnificent characteristics, although there have been challenges in 
implementing these in transparent electronics. Outstanding characteristics 
such as better mobility and retention have been observed in transparent 
ferroelectric memory devices which are composed of doped oxides, in turn 
exhibiting good stability\cite{wang}.

Indium oxide itself has a wide band gap of around 2.1 eV, which makes it a 
transparent material in the visible range that can be 
used in many areas for advantageous applications\cite{9}. On the other hand, 
gallium oxide has a band gap of 2.04 eV\cite{10,11}, which corresponds to the ultraviolet 
range. It is used in Schottky barrier diodes and field effect 
transistors, optoelectronic electroluminescent devices, spintronic devices and 
resistive random access memories, as a photocatalyst and many more\cite{23,24}. 
Indium oxide, a transparent semiconducting oxide, has been studied over the 
years and has been used as an active gas sensor material after detailed 
investigation of its various properties as a semiconductor\cite{12}. These 
have a 
wide band gap and when a vapor-deposited indium layer was oxidized at high 
temperatures in air, the resulting polycrystalline indium oxide was found 
to be transparent and conductive. Nanometer-sized crystalline indium oxide 
grains together form a polycrystalline film that is used today\cite{13}, and even 
monocrystalline nanowires are being researched. Its applications are also 
characterized by changing the band gap and forming heterostructures in the 
alloy system in combination with gallium oxide or even aluminum oxides. 

For advanced semiconductor applications, when indium oxide
is combined with another semiconducting oxide, a new path is made for bipolar 
devices with enhanced conductivity. Doping, tunable mobility and contact 
properties will act as the key requirements for further use \cite{14}.
Transparent conductive oxides are generally used as bold screens in 
optoelectronic devices, and in architectural and window glass. 
Composite oxides are used today in various optoelectronic applications such 
as photodetectors, optical modulators, polarizers, waveguides and lasers, 
as well as in batteries\cite{15}.

In this field of optoelectronics, researches are still exploring the 
various properties of these composite materials. The dynamical stability of this 
structure is determined using phonon density calculation. 
This paper aims to first 
geometrically optimize the heterostructure and then perform the energy bandgap 
and density of states calculations for this composite oxide of Indium oxide and 
Gallium oxide. An in-depth 
analysis of the optical properties has been carried out using density functional 
theory and optical constants. All of the optical properties including the 
refractive index, absorption and dielectric function are estimated.  
\section{Methodology}
The van der Waals heterostructure of the indium oxide and gallium oxide was 
formed 
theoretically, in which the layers of indium oxide are stacked over gallium 
oxide in (001) direction.  
Here, the ground state of the compound oxide containing indium oxide and 
gallium oxide is investigated for its optimized electronic structure (FIG.1) 
using 
density functional theory\cite{16}. The Tkatchenko-Scheffler method\cite{TS} 
(DFT-TS) with self-consistent screening was chosen to evaluate the vdW 
interactions and interlayer distance is chosen from minimum energy. 
Self-consistent field calculations are performed 
until a minimum energy value of 3.0x10$^{-5}$ eV/atom is reached. 
The crystal is structurally relaxed until the maximum force that can be exerted 
on it is less than 1.2 eV/$\AA$. The PBE(Perdew Burke Ernzerhof)+GGA(generalized 
gradient approximation) functional\cite{17} and the relativistic Koelling-Harmon 
treatment\cite{18} with an energy range of 3 eV. The band structure and density of 
states (DOS) are analyzed after the self-consistent field calculations have 
been performed. To structurally optimize the heterostructure, a fine k-set with $10\times10\times10$ k-mesh is used\cite{19}. Optical calculations of the 
imaginary and 
real parts of the dielectric tensor, reflectivity, conductivity, absorption 
and refractive index are performed using CASTEP software\cite{20}. The 
computational calculation of the phonon spectra is also carried out using 
norm-conserving pseudopotentials\cite{21} at a cut-off energy of 600eV. In the 
phonon spectra (FIG.2b), it is observed that density of states lies 
only in the positive energy range, indicating dynamical stablity of the material.
\section{Results}
The results of density functional theory are discussed in this section. This 
compound crystal has a P-1 symmetry with lattice 
parameters a=b=c 13.9981$\AA$ and $\alpha=\beta=\gamma=90^{\circ}$.
The atom's stability hinges on the self-consistent field-calculated energy, 
and we opt for the structure with the minimum energy to get insight into its 
other 
properties\cite{22}. 
In the band structure calculation, the k-points are taken according to the 
first Brillouin zone from $\Gamma$: (0.000 0.000 0.000) through 
F: (0.000 0.500 0.000), Q: (0.000 0.500 0.500), Z: (0.000 0.000 0.500), 
and back to $\Gamma$: (0.000 0.000 0.000) (FIG.2a). 
According to density functional 
theory, our results show a direct band gap of 3.097 eV.
\begin{figure}
\epsfig{file=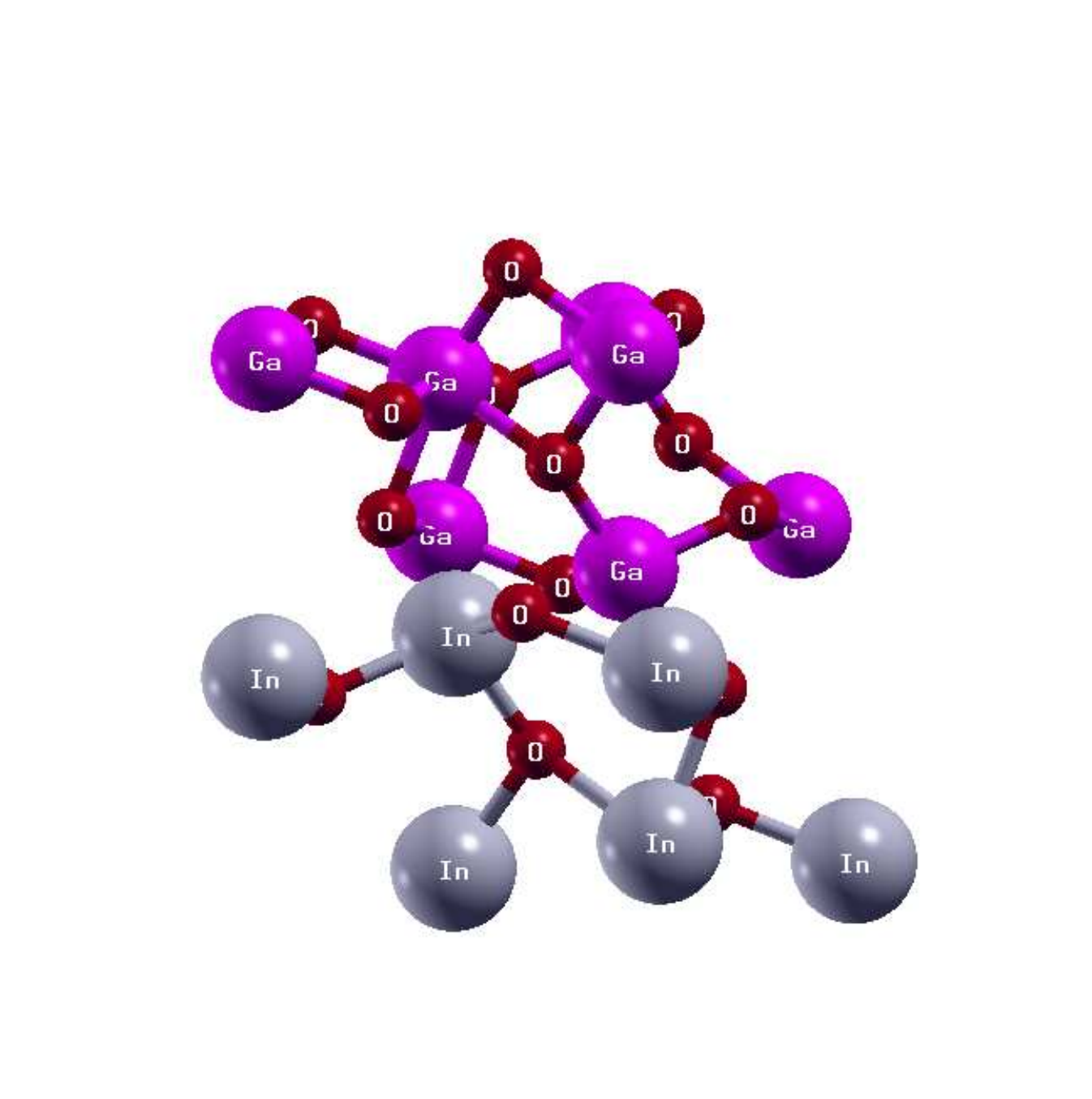,trim=2.0in 0.5in 0.5in 3.0in,clip=false, width=60mm}
\caption{(Color Online) Single unit crystal structure of layered indium oxide and gallium oxide.
	The atoms are marked and colored for reference.
}
\label{fig1}
\end{figure}
\begin{figure*}
\epsfig{file=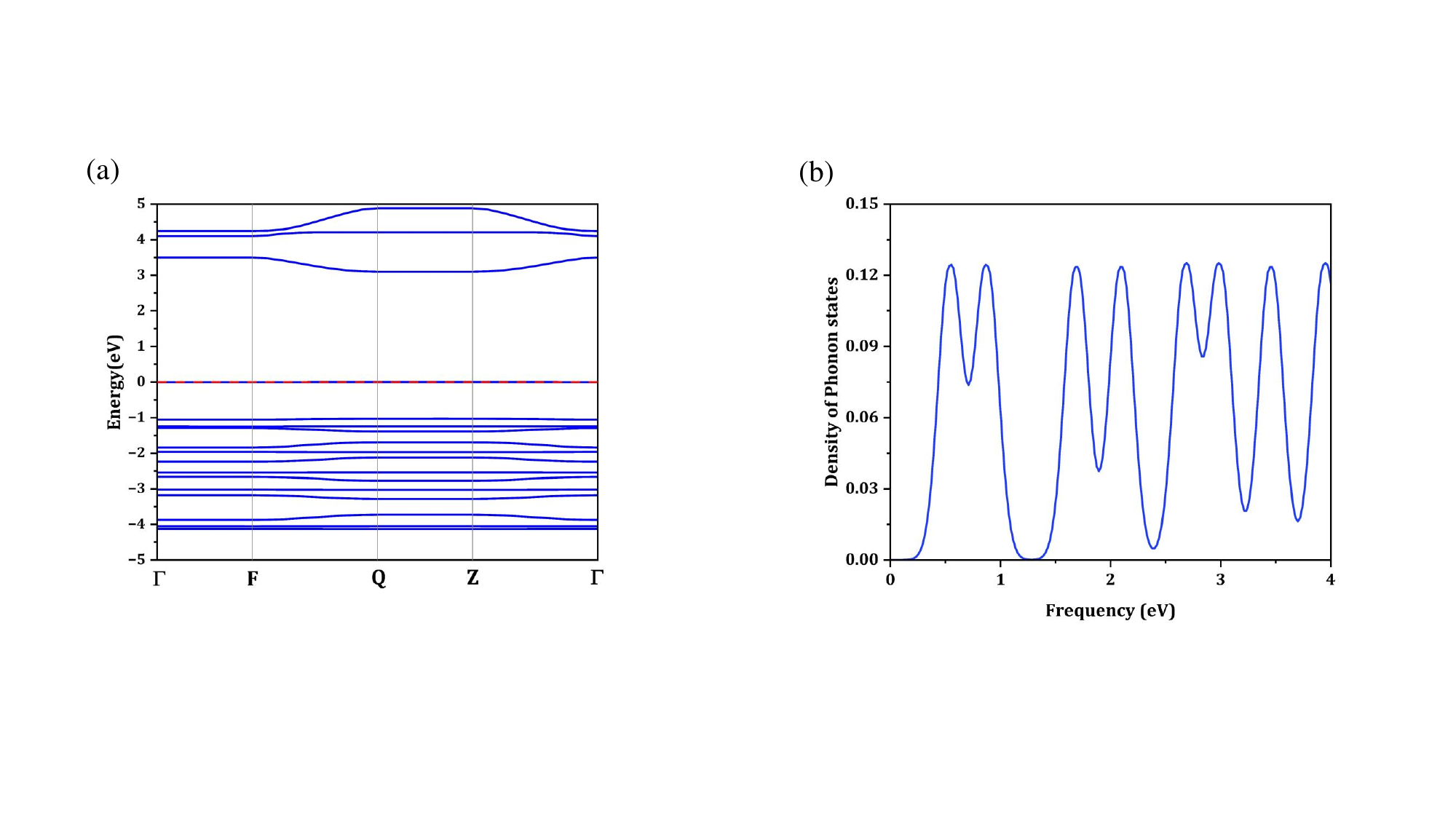,trim=0.8in 2.0in 1.8in 0.3in,clip=false, width=160mm}
	\caption{(Color Online) (a) Electronic band structure of layered oxide 
	which shows a band gap starting from Fermi Level. A 3 eV gap in the 
	band structure is a consistent phenomena in all the electronic 
	properties.(b) phonon density of states shows stability of the material. 
	}
\label{fig2}
\end{figure*}

\begin{figure*}
\epsfig{file=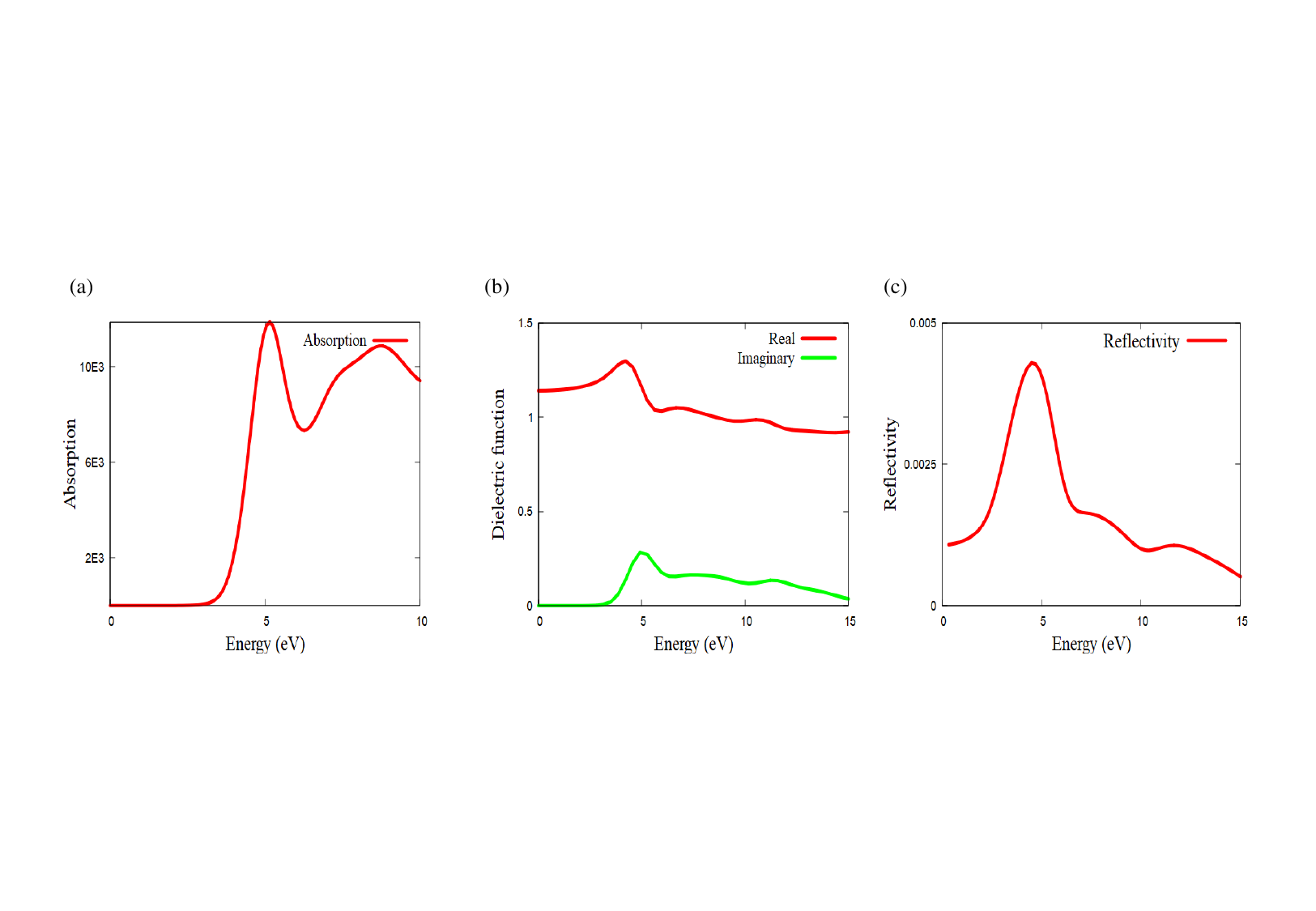,trim=1.5in 2.0in 1.5in 2.0in,clip=false, width=160mm}
	\caption{(Color Online) Plot of (a) absorption spectra, (b) frequency dependent 
	imaginary and real part of dielectric function and (c) reflectivity of the 
	composite oxide. In the dielectric function plot red line shows 
	real part and green line shows imaginary part. All the opto-electronic properties reveal a band gap of 3 eV }
\label{fig3}
\end{figure*}

A visual voyage into the band structure reveals a vivid picture of the 
electronic structure of a material, highlighting the valence band and conduction band, two key concepts in understanding basic physical properties. 
The valence band, spanning from -5 eV to 0 eV, represents the energy levels 
occupied by electrons that are tightly bound to their respective atoms. 
These electrons are involved in chemical bonding and are not free to move 
about the material with the illustrious Fermi level demarcated by a dotted line.
On the other side, the conduction band takes center stage on the right, 
showcasing its energy states from 0 eV to 5 eV. The band ranges from -5 eV to 5 eV
is considered because all the important phenomena are considered to be observed 
in this range.
The interplay between the valence band, conduction band, and Fermi level 
governs the electronic behavior of materials, shaping their electrical 
conductivity, optical properties, and thermal characteristics. Understanding 
these concepts 
is essential in fields such as semiconductor physics, materials science, and 
device engineering while proposing a new material.
The band gap found here is new and improved from its component layers. 
While the indium oxide shows a lower band gap around 2 eV and gallium oxide 
shows a band gap of about 5 eV the compound shows a gap of around 3 eV.
Our observations of direct band gap is important because it makes the 
material sufficiently bright and electron-hole radiative lifetime is small.

This section delves into the optical properties (FIG.3 and FIG.4) of the 
material, highlighting 
their potential applications in both the electronic and optical domains. 
The tetrahedron method is used for integrating over the Brillouin zone to calculate the optical properties.
The 
absorption spectra (FIG.3a) reveals a prominent peak in the visible spectrum, 
centered 
around 4eV. The first peak in the absorption spectra starts around 3 eV 
coinciding with the previously determined direct band gap of 
approximately 3.097 eV. This suggests a favorable absorption window where 
light in this range encounters minimal reflection, making these materials 
promising candidates for visible range light-emitting devices or photodetectors.

The real part of the dielectric function (FIG.3b) further unveils intriguing 
features. 
The static dielectric constant of the material is calculated from the value at 
zero frequency which is around 1.1 which gives a measure of the refractive index
of the material. Here we have plotted in-plane dielectric constant, however 
dielectric constant will show different behavior in the x-z direction proving 
the optical anisotropy of the material. In the dielectric function plot first 
there is a region of slow increase with energy followed by a peak around 4 eV. 
Then the dielectric function decreases with energy till 15 eV. This 
fascinating interplay of frequencies underscores the intricate relationship 
between optical and electronic properties in these materials.

These findings suggest promising avenues for exploration in the field of 
optoelectronics. The materials' ability to absorb and emit light in specific 
frequency ranges could lead to the development of novel light sources, 
detectors, and other optical components. Further research is warranted to 
fully unlock the potential of these materials and harness their unique 
optical properties for practical applications.

The intricate results of optics indeed reveal captivating insights into the 
material's properties. The refractive index, a measure of how light bends as 
it passes through a material, is observed from the refractive index vs energy
graph (FIG.4b). The intensity here reaches its 
peak between the infrared and visible spectrum, indicating a strong interaction 
of light with the material in this range. As we move towards the ultraviolet 
range, the refractive index gradually decreases, suggesting weaker  
interactions with higher-energy photons\cite{18}.
In the realm of visible light, the refractive index pirouettes around 1.1. 
This value is crucial for applications such as lens design and optical coatings,
as it determines how light gets refracted and transmitted.
Optical conductivity (FIG.4a), another key optical property, describes how 
a material 
responds to light by allowing the movement of carriers. The conductivity 
vs. energy graph reveals a peak around 5 eV, indicating a significant 
response to light in this energy range. This peak appears above the material's 
band gap energy, suggesting strong interactions between photons and electrons.
Meanwhile, the loss function (FIG.4c), which accounts for the energy absorbed 
by the 
material as light passes through, takes center stage with its own peak, 
belonging to the region with shorter wavelengths. This peak suggests that the 
material is more effective at absorbing higher-energy photons.

\begin{figure*}
\epsfig{file=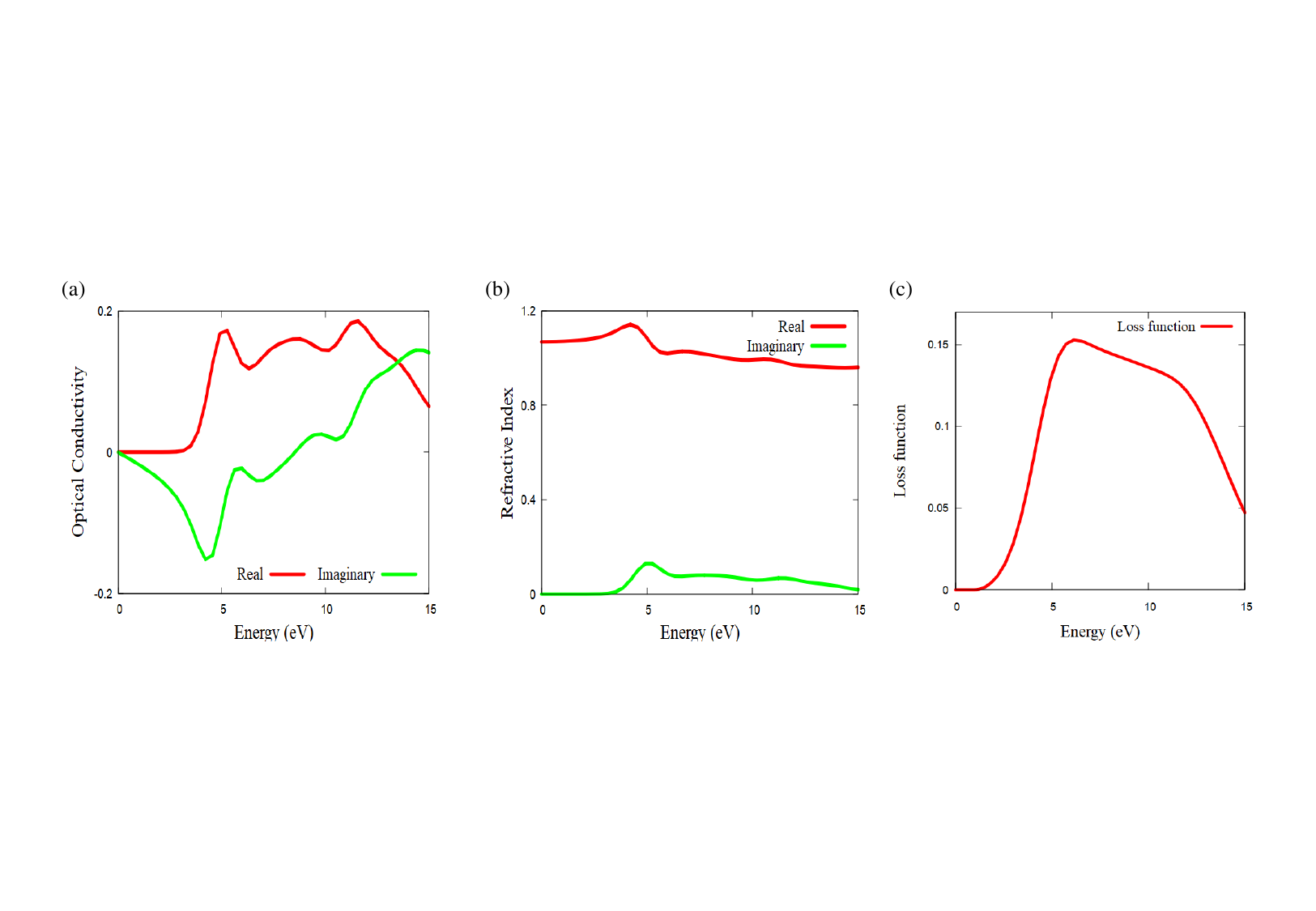,trim=1.5in 2.0in 1.5in 2.0in,clip=false, width=160mm}
	\caption{(Color Online) (a) Optical conductivity, (b) imaginary and real part of the refractive index (red line shows real part and green line 
	shows imaginary part)and (c) loss function is shown. }
\label{fig4}
\end{figure*}
\section{conclusion}
In conclusion, the composite crystal studied here unfurls a lot of 
possibilities in the opto-electronic fields. The union of Indium oxide and 
Gallium oxide in a heterostructure holds the promise of elevated 
optical prowess, and the results echo the potential for substantial 
contemporary value. Delving into the depths of dynamical stability adds another 
layer of comprehension to this crystal's stability.

The strategic use of a heterostructure emerges as a key player, and putting 
together all the results shows both electrical and optical properties, 
culminating in a commendable band gap energy value. This, in turn, 
opens doors to a plethora of applications, particularly in the field of 
solar cells and other opto-electronic devices.
The dielectric function and optical conductivity shows a peak after 3 eV 
proving the existence of a band gap.
The refractive index and the static value of dielectric constant is having a value of 1.1 which is a promising value for designing optical materials.

Armed with a favorable band gap nestled in the visible spectrum of the 
electromagnetic realm, these materials promise towards the construction of 
innovative opto-electronic devices. The meticulous 
exploration of this crystal's structure, guided by the results from the 
DFT method, adds a possible optoelectronic material made of two oxides.

\section{Data Availability}
The data that support the findings of this study are available from the corresponding author upon reasonable request.
	

\begin{thebibliography}{56}
\bibitem{1}E. Roduner, Size matters: why nanomaterials are different. 
	{\it Chem. Soc. Rev.}, {\bf 35}, 583 (2006); A.M. Smith, S. Nie, Semiconductor Nanocrystals: Structure, Properties, and Band Gap Engineering. {\it Acc. Chem. Res.}, {\bf 43}, 190 (2010).
\bibitem{2}Wang, Xinzhuo, The Comparison of Titanium Dioxide and Zinc Oxide Used in Sunscreen Based on Their Enhanced Absorption. {\it  Applied and Computational Engineering.}, {\bf 24}, 237 (2023).
\bibitem{3}Choi Y-S, Kang J-W, Hwang D-K , Park S-J, Recent advances in ZnO-based light emitting diodes. {\it IEEE Trans. Electr. Dev.}, {\bf 57}, 26 (2010).
\bibitem{4}V.M. Bermudez, The structure of low-index surfaces of $\beta$-Ga$_2$O$_3$. {\it Chemical Physics}, {\bf 323}, 193  (2006).
\bibitem{5}Q.H. Wang, et al., Electronics and optoelectronics of two-dimensional transition metal dichalcogenides. {\it Nat. Nanotechnol.}, {\bf 7}, 699 (2012).
\bibitem{6}J. Pascual, J. Camassel, H. Mathieu, Fine structure in the intrinsic absorption edge of TiO$_2$. {\it Phys. Rev. B.}, {\bf 18}, 5605 (1978).
\bibitem{7}H. Tang, F. L´evy, H. Berger, P.E. Schmid, Urbach tail of anatase TiO$_2$. {\it Phys. Rev. B.}, {\bf 52}, 7771 (1995).
\bibitem{8}Juliya Khanam, Md. Rashib Hasan, Bristy Biswas, Shirin Akter Jahan, Nahid Sharmin, Samina Ahmed, Sharif Md. Al-Reza, Development of ceramic grade red iron oxide pigment from waste iron source, {\it Heliyon.}, {\bf 9}, e12854 (2023).

\bibitem{wang}Wang, Z., Nayak, P. K., Caraveo-Frescas, J. A., and Alshareef, H. N. Recent Developments in p-Type Oxide Semiconductor Materials and Devices.{\it Advanced Materials}, {\bf  28}, 3831-3892 (2016).
\bibitem{9}Ma, Zhizhen, Li, Zhuoran, Liu, Ke, Ye, Chenran and Sorger, Volker J, Indium-Tin-Oxide for High-performance Electro-optic Modulation. {\it Nanophotonics.},{\bf 4}, 198 (2015); Hillier, J.A., Patsalas, P., Karfaridis, D. et al., Photo-engineered optoelectronic properties of indium tin oxide via reactive laser annealing. {\it Sci Rep 12.}, {\bf 12}, 14986 (2022).
\bibitem{10}Jamwal NS, Kiani A, Gallium Oxide Nanostructures: A Review of Synthesis, Properties and Applications, {\it  Nanomaterials (Basel).}, {\bf 12}, 2061 (2022).
\bibitem{11}D.Guo,Q. Guo, Z. Chen, Z. Wu, P. Li, W. Tang, Review of Ga$_2$O$_3$-based optoelectronic devices. {\it Materials Today Physics.}, {\bf 11}, 100157 (2019).

\bibitem{23}Oreibi, I., Habeeb, M.A. and Hamza, R.S.A, Tailoring the 
	Structural and Optical Features of PVA/SiO$_2$-CuO Polymeric 
		Nanocomposite for Optical and Gamma Ray Shielding 
		Applications. {\it Silicon}, 1-13 (2023).
\bibitem{24}Hashim A, Habeeb MA, Jebur QM, Structural, dielectric and optical properties for (Polyvinyl alcohol-polyethylene oxide manganese oxide) nanocomposites. {\it Egypt J Chem}, {\bf 63}, 735-749 (2020).
\bibitem{12}Hadeel Salih Mahdi, Azra Parveen, Mawlood Maajal Ali, Ameer Azam, Microstructural and Optical properties of Indium Oxide Nanoparticles. {\it Materials Today: Proceedings.},{\bf 18}, 704 (2019).
\bibitem{13}Lee, Jong Hoon, Kim, Young, Ahn, Sang, Ha, Tae, Kim, Hong Seung, Grain-size effect on the electrical properties of nanocrystalline indium tin oxide thin films. {\it Materials Science and Engineering.}, {\bf 199}, 37 (2015).
\bibitem{14}Bierwagen, Oliver, Indium oxide - A transparent, wide-band gap semiconductor for (opto)electronic applications. {\it Semiconductor Science and Technology.}, {\bf 30}, 024001 (2015).
\bibitem{15}Nicolas Barreau, David Duche, Carmen M. Ruiz, Ludovic Escoubas, Jean-Jacques Simon, Judikael Le Rouzo, Veronica Bermudez, Chapter 16 - Innovative approaches in thin-film photovoltaic cells. {\it Woodhead Publishing}, 595 (2018).
\bibitem{16}Nidhin Kurian Kalarickal, Siddharth Rajan, Chapter Three - $\beta$-
	(Al$_x$Ga$_{(1-x)})$$_2$O$_3$ epitaxial growth, doping and transport, Editor(s): Yuji Zhao, Zetian Mi, Semiconductors and Semimetals,{\it Elsevier.}, {\bf 107}, 49 (2021).
\bibitem{TS}A. Tkatchenko and M. Scheffler, Accurate Molecular Van Der Waals Interactions from Ground-State Electron Density and Free-Atom Reference Data, {\it Phys. Rev. Lett.}, {\bf 102}, 073005 (2009).		
\bibitem{17}Xiaoping Han and Noureddine Amrane and Maamar Benkraouda, A GGA + vdW study on electronic properties and optoelectronic functionality of Cd-doped tetragonal CH$_3$NH$_3$PbI$_3$ for photovoltaics. {\it Chemical Physics}, {\bf 556}, 111461 (2022).
\bibitem{18}Koelling, Dale Harmon, Bruce, A technique for relativistic spin-polarised calculations. {\it Journal of Physics C: Solid State Physics.}, {\bf 10}, 3107 (2001).
\bibitem{19}S. Koley, Saurabh Basu, Superconductivity induced by Ag intercalation in Dirac semimetal Bi2Se3. {\it Computational Materials Science.}, {\bf 210},   110989 (2022).
\bibitem{20}S. Koley, Engineering Si doping in anatase and rutile TiO$_2$ with oxygen vacancy for efficient optical application. {\it Physica B: Condensed Matter.}, {\bf 602}, 412502 (2021).
\bibitem{21}D. R. Hamann, M. Schlüter, and C. Chiang, Norm-Conserving Pseudopotentials. {\it Phys. Rev. Lett.}, {\bf  43}, 1494 (1979).
\bibitem{22}S. Koley, Theoretical study on spintronic and optical property prediction of doped magnetic Borophene, {\it Computational Condensed Matter.}, {\bf 34}, e00783 (2023).

\end{thebibliography}
\end{document}